\documentclass [11pt,a4paper] {article}

\usepackage[cp1252]{inputenc}

\usepackage{amssymb}
\usepackage{amsmath}
\usepackage{amsfonts,amssymb}

\usepackage[dvips]{graphicx}

\DeclareMathAlphabet{\mathpzc}{OT1}{pzc}{m}{it}

\begin{document}

\title{Examples of non-constructive proofs in quantum theory}

\author{Arkady Bolotin\footnote{$Email: arkadyv@bgu.ac.il$} \\ \textit{Ben-Gurion University of the Negev, Beersheba (Israel)}}

\maketitle

\begin{abstract}\noindent Unlike mathematics, in which the notion of truth might be abstract, in physics, the emphasis must be placed on algorithmic procedures for obtaining numerical results subject to the experimental verifiability. For, a physical science is exactly that: algorithmic procedures (built on a certain mathematical formalism) for obtaining verifiable conclusions from a set of basic hypotheses. By admitting non-constructivist statements a physical theory loses its concrete applicability and thus verifiability of its predictions. Accordingly, the requirement of constructivism must be indispensable to any physical theory. Nevertheless, in at least some physical theories, and especially in quantum mechanics, one can find examples of non-constructive statements. The present paper demonstrates a couple of such examples dealing with macroscopic quantum states (i.e., with the applicability of the standard quantum formalism to macroscopic systems). As it is shown, in these examples the proofs of the existence of macroscopic quantum states are based on logical principles allowing one to decide the truth of predicates over an infinite number of things.\\

\end{abstract}

\section{Introduction: The measurement paradox in quantum mechanics}\label{Introduction}

\noindent As it is known, originally the quantum formalism was only designed to explain phenomena occurring in the region of single electrons, photons (i.e., the photoelectric effect) and atoms. But since quantum mechanics was introduced, this region has been gradually extended and at the present time length scales over which quantum mechanics has been directly tested (in the sense of detection of characteristically quantum effects such as interference) are down to about $10^{-18}$ m in high-energy diffraction experiments \cite{Leggett, Schlosshauer06} and up to one picometer ($10^{-12}$ m) in modern interferometric experiments demonstrating the quantum wave nature of large organic molecules (composed of up to 430 atoms, with a maximal size of up to 60 \AA) \cite{Gerlich} and as far as 143 km in free-space entanglement-swapping experiments, verifying the presence of quantum entanglement between two previously independent photons \cite{Zeilinger}.\\

\noindent Even though this region (where quantum mechanics has been directly tested and presented satisfactory and often excellent evidence that the quantum formalism is quantitatively valid) continues to be rather limited, the attitude of the great majority of practicing physicists is to assume \textit{the universal validity of the quantum formalism} over the whole scale of distance \cite{Schlosshauer04,Hajicek}. The logic behind such an assumption can be explained as follows:\\

\noindent In compliance with the principle of excluded middle, \textit{the quantum formalism can be either valid or not}; so, since there is no known experimental evidence that the quantum formalism is not valid in the regions, where quantum mechanics has not been directly tested, this formalism must be valid for the whole of the physical universe without restriction.\\

\noindent What is most importantly is that the assumption of the universal validity of the quantum formalism does not supply a method (i.e., algorithm) to exhibit explicitly a wave function (or a quantum state vector) for any arbitrary physical system. Specifically, the general form of the time-dependent Schr\"odinger equation (presented in most interpretations of quantum mechanics) provides no indication as to how to construct (that is, calculate) the wave function whose existence is claimed by the quantum formalism.\\

\noindent Together with all that, the application of the quantum formalism to macroscopic systems appears to lead immediately to a conflict with everyone’s experience of the everyday world, known as \textit{the quantum measurement paradox}. By way of explanation, if the quantum formalism is taken to apply to all physical systems and hence a wave function exists for any physical system, including a macroscopic measurement apparatus, then when that apparatus interacts with another system, the joint state should be a superposition of wave functions; and yet, one never observes such a macroscopic quantum superposition. For that reason, most of the papers concerning with the foundation of quantum formalism concentrate on resolving this famous paradox \cite{Tammaro}.\\

\noindent However, rather than trying to find a solution to the given paradox, it is also possible \textit{to not assume the principle of excluded middle as an axiom} (causing this paradox in the first place). More explicitly, it is possible to show that quantum states of an arbitrary macroscopic system cannot be constructive (i.e., effectively calculable), and thus their existence cannot be proved (verified) \textit{with any algorithmic procedure} for obtaining a conclusion from the set of fundamental hypotheses of the quantum formalism.\\

\noindent With such a goal in mind, in this paper, we will give examples demonstrating that the application of the quantum formalism to at least some macroscopic systems implies principles that are known to be nonconstructive (that is, uncomputable).\\

\section{Finding the global minimum for the energy of the quantum Ising model}\label{Ising_model}

\noindent Going with the paper \cite{Bolotin1}, let us consider the quantum version of the classical infinite-range antiferromagnetic Ising model of the spin glass controlled by the following Hamiltonian\\

\begin{equation} \label{Ising_Hamiltonian} 
     H\!\left(\!  \sigma_{1}^{z},\sigma_{2}^{z}, \dotsc, \sigma_{N}^{z}  \!\right)
      =
      \left(
      \sum_{i=1}^{N}q_i \sigma_{i}^{z}
      \right)^2
      \;\;\;\;   ,
\end{equation}
\smallskip

\noindent where $q_i$ are some positive numbers of arbitrary size and each of $N$ discrete variables (“spins”) $S_{ik}=\pm1$ of the classical Ising model is replaced by the Pauli matrix $\sigma_{i}^{z}$ acting on the $i^{\mathrm{th}}$ qubit labeled by $\left|\!\left.{z_{ik}} \!\right.\right\rangle$ with $z_{ik} \in\{-1,+1\}$ such that $S_{ik}=+1$ corresponds to $\left|\!\left.{z_{ik}=+1} \!\right.\right\rangle$ (i.e., the $i^{\mathrm{th}}$ quantum spin being up in the $z$-direction) and $S_{ik}=-1$ corresponds to $\left|\!\left.{z_{ik}=-1} \!\right.\right\rangle$ (i.e., the $i^{\mathrm{th}}$ quantum spin down in the $z$-direction). As it can be readily seen, the Hilbert space of the considered quantum model is spanned by the $2^N$ basis vectors $\left|\!\left.{z_{1k}} \!\right.\right\rangle, \left|\!\left.{z_{2k}} \!\right.\right\rangle, \dotsc, \left|\!\left.{z_{Nk}} \!\right.\right\rangle$ expressing all possible spin configurations of the given quantum model.\\

\noindent Let $\left|\!\left.{\psi_{k}} \!\right.\right\rangle = \left|\!\left.{z_{1k}} \!\right.\right\rangle \left|\!\left.{z_{2k}} \!\right.\right\rangle \cdots \left|\!\left.{z_{Nk}} \!\right.\right\rangle$ be the  $k^{\mathrm{th}}$ spin configuration of this model, that is, the $k^{\mathrm{th}}$ eigenstate of the Hamiltonian (\ref{Ising_Hamiltonian}). Also, let $A_k$ be the subset of the index set $I=\{1,2,\dotsc,N\}$ which identify the spins in the configuration $\left|\!\left.{\psi_{k}} \!\right.\right\rangle$ being up in the $z$-direction:\\

\begin{equation} \label{Subset_A} 
     A_k
      =
      \big\{
      i \in {I} \,\big |\,
                                   \left|\!\left.{z_{ik}=+1} \!\right.\right\rangle  \!
      \big\}
      \;\;\;\;   .
\end{equation}
\smallskip

\noindent Then, as it is readily apparent, every eigenvalue $E_k$ of the Hamiltonian (\ref{Ising_Hamiltonian}) can be presented as the outputs of the function $E(A_k)$ defined by the set of the subsets $A_k$\\

\begin{equation} \label{E_k} 
     E_k
      =
     E\!\left(A_k\right)
      =
      \Big(
      \sum_{i \in A_k}q_i \; - \sum_{i \in I\setminus A_k}\! q_i
      \Big)^2
      \;\;\;\;   .
\end{equation}
\smallskip

\noindent Accordingly, the problem of finding the ground energy $E_{\mathrm{ground}}$ of the quantum Ising model\\

\begin{equation} \label{SE} 
      H\!\left(\!  \sigma_{1}^{z},\sigma_{2}^{z}, \dotsc, \sigma_{N}^{z}  \!\right)
                                                                                                                                               \left|\!\left.{\psi_{\mathrm{ground}}} \!\right.\right\rangle
      =
      E_{\mathrm{ground}}
                                                                                                                                               \left|\!\left.{\psi_{\mathrm{ground}}} \!\right.\right\rangle
      \;\;\;\;    
\end{equation}
\smallskip

\noindent would be equivalent to the problem of finding such a subset  $A_k=A_{\mathrm{ground}}$ that minimize the function $E (A_k)$, i.e.,\\

\begin{equation} \label{Min_energy} 
     E_{\mathrm{ground}}
      =
      \min {\Big\{E\!\left(A_k\right)\!\Big\}_{k=1}^{2^N}}
      \;\;\;\;   .
\end{equation}
\smallskip

\noindent Now, if the quantum formalism is applicable to macroscopic systems in the same way as it is to microscopic systems, then in \textit{the macroscopic limit}\footnote{It is a limit for a large number of system's constituent particles $N$, such as atoms or molecules, where the size of the system is taken to grow in proportion with $N$. In the macroscopic limit $N \rightarrow \infty$, which is a special case of the limit $\hbar\rightarrow 0$, microscopic systems are considered finite, whereas macroscopic systems are infinite \cite{Landsman}.} the expression (\ref{Min_energy}) must give the global minimum for the macroscopic energy of the classical Ising model (that is, the lowest possible value of the ensemble average $\langle E \rangle =\sum_k{H_{k}P_{k}}$ for the energy, which is the sum of the microstate energies $H_k:=H(S_{1k},S_{2k},\dotsc, S_{Nk})$ weighted by their probabilities $P_{k}$\\

\begin{equation} \label{Probabilities} 
     P_{k}
      =
      \frac{1}{Z} e^{-\beta H_{k}}
      \;\;\;\;   ,
\end{equation}
\smallskip

\noindent where $\beta$ is the thermodynamic beta and $Z$ is the partition function defined over all possible spin configurations of the classical model):\\

\begin{equation} \label{Macroscopic_energy} 
      \lim_{N \rightarrow \infty} \min {\Big\{E\!\left(A_k\right)\!\Big\}_{k=1}^{2^{N}}}
     =
     \min \!{\langle E \rangle}
     \;\;\;\;   .
\end{equation}
\smallskip

\noindent Inasmuch as the function (\ref{E_k}) is bounded from below, this global minimum $\min \!{\langle E \rangle}$ does exist conforming to classical analysis. However, the fact that there exists a global minimum of the function $E\!\left(A_k\right)$ (and hence, in principle, it is possible to find it) does not mean that one can actually find it!\\

\noindent Obviously, the problem is that even if an actual (non–zero) global minimum of $E\!\left(A_k\right)$ has been just reached, there is no way (i.e., an effective algorithm) \textit{to recognize it as such}. Indeed, defining the property of being the ground energy $E_{\mathrm{ground}}$ in the macroscopic limit ${N \rightarrow \infty}$ involves quantification over the set of all subsets $A_k$ that has the cardinality of the continuum\footnote{In the limit ${N \rightarrow \infty}$ the index set $I=\{1,2,\dotsc,N\}$ becomes the set $\mathbb{N}$ of the natural numbers and accordingly the set of all subsets $A_k$ of $I$ becomes the power set $\mathcal{P}(\mathbb{N})$.}. Consequently, to recognize $\min \!{\langle E \rangle}$ as the global minimum of the function $E\!\left(A_k\right)$ would require one to check the inequality $\min \!{\langle E \rangle} \le E\!\left(A_k\right)$ \textit{uncountably many} times.\\

\noindent This implies that one cannot actually find (construct) the eigenvector $\left|\!\left.{\psi_{\mathrm{ground}}} \!\right.\right\rangle$ (or the set of the eigenvectors $\left|\!\left.{\psi_{\mathrm{ground}}} \!\right.\right\rangle$) corresponding to the global minimum $\min \!{\langle E \rangle}$ and therefore demonstrate the existence of a macroscopic quantum superposition for the Ising model (\ref{Ising_Hamiltonian}).\\

\section{Recognizing Fock states of the quantum model that are associated with Fibonacci numbers}\label{Fibonacci_states}

\noindent Analogous to the paper \cite{Bolotin2}, let us consider the quantum model of $N$ non-interacting identical particles whose Hamiltonian mimics the form of the left–hand–side squared of the Diophantine equation for non-negative integers:

\begin{equation} \label{Diophantine_Hamiltonian}
      H=\left(
                                  a^{\dagger}_{3}a_{3}
                                - a^{\dagger}_{2}a_{2}
                                - a^{\dagger}_{1}a_{1}
                     \right)^2
      \;\;\;\;  ,
\end{equation}
\smallskip

\noindent where the creation $a^{\dagger}_j$ and annihilation $a_j$ operators (similar to those of the three-dimensional quantum harmonic oscillator) together form the number operators $\hat{N_j} {:=}a^{\dagger}_{j}a_j$ which have only non-negative integer eigenvalues $n_j$ and whose eigenstates $\left|\!\left.{\psi} \!\right.\right\rangle$ are those of the Hamiltonian (\ref{Diophantine_Hamiltonian})\\

\begin{equation} \label{Eigenstates} 
     \begin{array}{l l}
         \hat{N_j}\left|\!\left.{\psi} \!\right.\right\rangle = n_j \! \left|\!\left.{\psi} \!\right.\right\rangle                                                                                                              \;\;\;\;   , \\
         H\!\left|\!\left.{\psi} \!\right.\right\rangle = \left(n_1-n_2-n_3\right)^2 \left|\!\left.{\psi} \!\right.\right\rangle {:=} E\! \left|\!\left.{\psi} \!\right.\right\rangle\;\;\;\;   .
      \end{array}
      \;\;\;\;   
\end{equation}
\smallskip

\noindent As it can be readily seen, the zero ground state $\left|\!\left.{\psi_{0}} \!\right.\right\rangle$ of the Hamiltonian (\ref{Diophantine_Hamiltonian}) (that is, the state with the zero ground energy $E=0$) will be a linear superposition of Fock states (that is, states with definite particle number \cite{Altland})

\begin{equation} \label{Superposition} 
       \left|\!\left.{\psi_{0}} \!\right.\right\rangle
        =
        \sum_{i} c_i 
                                                \left|\!\left.{n_{1_i}} \!\right.\right\rangle\!
                                                \left|\!\left.{n_{2_i}} \!\right.\right\rangle\!
                                                \left|\!\left.{n_{3_i}} \!\right.\right\rangle
      \;\;\;\;  ,
\end{equation}
\smallskip

\noindent where $n_{j_i}$ specifies the number of particles in the $i^{\mathrm{th}}$ state $j_i$ of the model. Those numbers meets the obvious condition $\sum_{j}^{3}\sum_{i}n_{j_i}=N$, while the superposition coefficients $c_i$ meet the normalization requirement $\sum_{i}|c_i|^2=1$.\\

\noindent Among the non-vacuum states $\left|\!\left.{n_{1_i}} \!\right.\right\rangle\!\left|\!\left.{n_{2_i}} \!\right.\right\rangle\!\left|\!\left.{n_{3_i}} \!\right.\right\rangle$ (with nonzero number of particles) comprising the superposition (\ref{Superposition}) one may find \textit{Fibonacci states}  $\left|\!\left.{F_{1_i}} \!\right.\right\rangle\!\left|\!\left.{F_{2_i}} \!\right.\right\rangle\!\left|\!\left.{F_{3_i}} \!\right.\right\rangle$, that is, such states that\\

\begin{equation} \label{Fibonacci_states} 
     \begin{array}{l l}
         n_{1_i}=F_{1_i} \;\;\;\;   , \\
         n_{2_i}=F_{2_i} \;\;\;\;   , \\
         n_{3_i}=F_{3_i} \;\;\;\;   , 
      \end{array}
      \;\;\;\;   
\end{equation}
\smallskip

\noindent where $F_{1_i}$, $F_{2_i}$, and $F_{3_i}$ are sequential Fibonacci numbers connected by the recursion relation $ F_{3_i}=F_{1_i}+F_{2_i}$ (in the vacuum state $\left|\!\left.{0_{1_i}} \!\right.\right\rangle\!\left|\!\left.{0_{2_i}} \!\right.\right\rangle\!\left|\!\left.{0_{3_i}} \!\right.\right\rangle$ all $F_{j_i}=0$).\\

\noindent Since the set of natural numbers $\mathbb{N}$ can be written as the direct sum $\mathbb{N}=F \oplus Z$ of two of its proper subsets, the Fibonacci $F$ and non-Fibonacci $Z$ numbers, the eigenspace $\mathcal{E}_0$ of the zero ground energy $E=0$ for the considered quantum model can be expressible as the direct sum of two subsets $\mathcal{E}_F$ and $\mathcal{E}_Z$ formed by the Fibonacci and non-Fibonacci states, respectively,\\

\begin{equation} \label{Eigenspace} 
       \mathcal{E}_0 = \mathcal{E}_F \oplus \mathcal{E}_Z
        =
       \{ \left|\!\left.{F_{1_i}} \!\right.\right\rangle\!\left|\!\left.{F_{2_i}} \!\right.\right\rangle\!\left|\!\left.{F_{3_i}} \!\right.\right\rangle \}
       \oplus
       \{ \left|\!\left.{z_{1_i}} \!\right.\right\rangle\!\left|\!\left.{z_{2_i}} \!\right.\right\rangle\!\left|\!\left.{z_{3_i}} \!\right.\right\rangle \}
      \;\;\;\;  ,
\end{equation}
\smallskip

\noindent where the non-Fibonacci states are defined by\\

\begin{equation} \label{non_Fibonacci} 
       \left|\!\left.{z_{1_i}} \!\right.\right\rangle\!\left|\!\left.{z_{2_i}} \!\right.\right\rangle\!\left|\!\left.{z_{3_i}} \!\right.\right\rangle
        \in
       \{ \left|\!\left.{n_{1_i}} \!\right.\right\rangle\!\left|\!\left.{n_{2_i}} \!\right.\right\rangle\!\left|\!\left.{n_{3_i}} \!\right.\right\rangle \}
       \backslash
       \{ \left|\!\left.{F_{1_i}} \!\right.\right\rangle\!\left|\!\left.{F_{2_i}} \!\right.\right\rangle\!\left|\!\left.{F_{3_i}} \!\right.\right\rangle \}
      \;\;\;\;   
\end{equation}
\smallskip

\noindent and the vacuum state $\left|\!\left.{0_{1_i}} \!\right.\right\rangle\!\left|\!\left.{0_{2_i}} \!\right.\right\rangle\!\left|\!\left.{0_{3_i}} \!\right.\right\rangle$ belongs to the intersection $\mathcal{E}_F \cap \mathcal{E}_Z$. Subsequently, the zero ground state $\left|\!\left.{\psi_{0}} \!\right.\right\rangle$ can be extended as the superposition of the Fibonacci and non-Fibonacci states\\

\begin{equation} \label{Superposition2} 
       \left|\!\left.{\psi_{0}} \!\right.\right\rangle
        =
        c_i \left|\!\left.{0_{1_i}} \!\right.\right\rangle\!\left|\!\left.{0_{2_i}} \!\right.\right\rangle\!\left|\!\left.{0_{3_i}} \!\right.\right\rangle
        +
        \sum_{k} {\alpha}_k  \left|\!\left.{F_{1_k}} \!\right.\right\rangle\!\left|\!\left.{F_{2_k}} \!\right.\right\rangle\!\left|\!\left.{F_{3_k}} \!\right.\right\rangle
        +
        \sum_{l} {\beta}_l  \left|\!\left.{z_{1_l}} \!\right.\right\rangle\!\left|\!\left.{z_{2_l}} \!\right.\right\rangle\!\left|\!\left.{z_{3_l}} \!\right.\right\rangle
      \;\;\;\;   .
\end{equation}
\smallskip

\noindent If the quantum formalism is universally valid, this superposition should exist for any amount of particles $N$ including the case of the macroscopic model with ${N \rightarrow \infty}$.\\

\noindent It is clear that in order to constructively prove the existence of a superposition containing the Fibonacci and non-Fibonacci states is necessary to demonstrate a finite procedure able to decide whether or not a positive integer $n_{j_i}$ is a Fibonacci number. The problem, however, is that it is impossible to provide such a finite procedure in the macroscopic limit ${N \rightarrow \infty}$. Indeed, to recognize the Fibonacci numbers, one can apply either a straightforward (brute-force) procedure or the closed-form expression for the Fibonacci numbers\footnote{This closed-form expression is known as Binet’s formula \cite{Seroul}.}; but, whichever method is used, to reach the answer (whether $n_{j_i}$ is a Fibonacci number) in the macroscopic limit ${N \rightarrow \infty}$ would require an infinite amount of steps.\\

\noindent A brute-force procedure will generate the Fibonacci numbers until one becomes equal to a given positive integer $n_{j_i}$: If it does, then $n_{j_i}$ is a Fibonacci number, if not, the numbers will eventually become bigger than $n_{j_i}$, and the procedure will stop. But if the integer $n_{j_i}$ is unlimited – as it can be in the case of the considered macroscopic model with an infinite number of non-interacting identical particles – then the brute-force procedure would never terminate.\\

\noindent Based on the closed-form expression for the Fibonacci numbers, the positive integer  $n_{j_i}$ would belong to the Fibonacci sequence if the following equality holds:\\

\begin{equation} \label{Equality} 
       \bigg\lfloor
            n_{j_i} \!\left(1+\frac{p_n}{q_n}\right) + \frac{1}{n_{j_i}}
        \bigg\rfloor
        =
       \bigg\lceil
            n_{j_i} \!\left(1+\frac{p_n}{q_n}\right) - \frac{1}{n_{j_i}}
        \bigg\rceil
      \;\;\;\;  ,
\end{equation}
\smallskip

\noindent where $\lfloor\cdot\rfloor$ and $\lceil\cdot\rceil$ stand for the floor and ceiling functions, respectively, and the expression in parentheses is the golden ratio $\phi$ calculated to the accuracy of the  $n^\mathrm{th}$  Diophantine approximation\\

\begin{equation} \label{Golden_ratio} 
       \phi
        =
        \frac{1}{2} \left( 1 + \sqrt{5} \right)
        \cong 
        \big{\lbrack} {1;\underbrace{1,1,1,\dots,1}_n} \,\big{\rbrack}
        =
        1+\frac{p_n}{q_n}
      \;\;\;\;  
\end{equation}
\smallskip

\noindent such that $p_n$ and $q_n$ are given by the recurrence relation\\

\begin{equation} \label{Recurrence_relation} 
     \begin{array}{l l}
         p_n = q_{n-1} \;\;\;\;   , \\
         q_n =q_{n-1} + q_{n-2} \;\;\;\;    
      \end{array}
      \;\;\;\;   
\end{equation}
\smallskip

\noindent with the seed values $p_1 =1$ and $q_1 =1$. So, to correctly decide whether $n_{j_i}$ is a Fibonacci number, the upper bound for the Diophantine approximations ${p_n}/{q_n}$  of the infinite continued fraction $\{\phi\}=\phi-1$ given by the expression \cite{Waldschmidt}\\

\begin{equation} \label{Upper_bound} 
      \bigg|
      \{\phi\} - \frac{p_n}{q_n} 
      \bigg|
      <
      \frac{1}{\sqrt{5}{q_n}^2}
      \;\;\;\;   
\end{equation}
\smallskip

\noindent must be much less than the reciprocal of the integer  $n_{j_i}$, meaning that the golden ratio $\phi$ must be calculated to such an accuracy that the following inequality holds\\

\begin{equation} \label{Inequality} 
      {q_n}^2
      \gg
      \frac{n_{j_i}}{\sqrt{5}} 
      \;\;\;\;  .
\end{equation}
\smallskip

\noindent In the macroscopic limit ${N \rightarrow \infty}$ the integer  $n_{j_i}$ may be unlimited; accordingly, it would be necessary to calculate the fraction  $\{\phi\}$ to an unbounded accuracy ${p_{\infty}}/{q_{\infty}}$, which certainly could be achieved only by way of applying the recurrence relation (\ref{Recurrence_relation}) infinitely many times.\\

\noindent Thus, for the considered quantum model, the existence of a macroscopic superposition containing the Fibonacci and non-Fibonacci states (and in this way the validity of the quantum formalism in the macroscopic limit ${N \rightarrow \infty}$) cannot be constructively proven.\\

\section{A "weak" example of non-constructivism of the quantum formalism}\label{Weak_example}

\noindent Let us also give a "weak" example of non-constructivism of the quantum formalism. Such an example will not disprove the universal validity of the quantum formalism, but it will show that, at present, no constructive proof of its applicability to a particular macroscopic system is known.\\

\noindent Our weak example begins by taking the considered in the Section \ref{Ising_model} quantum Ising model in a more `realistic' macroscopic limit, such as  $N \rightarrow N_{\!A}$, where $N_{\!A} \sim 10^{24}$ is Avogadro's number.\\

\noindent On the one hand, because in the limit $N \rightarrow N_{\!A}$ the set of all subsets $A_k$ is bounded, to recognize $\min \!{\langle E \rangle}$ as the global minimum of the function $E\!\left(A_k\right)$ will now require a finite number of steps.\\

\noindent But, on the other hand, because the Hamiltonian of the quantum Ising model (\ref{Ising_Hamiltonian}) is NP-hard, recognizing $\min \!{\langle E \rangle}$ as the global minimum of $E\!\left(A_k\right)$ in the worst case might take $O(2^{N_{\!A}/2})$ amount of time and $O(2^{N_{\!A}/4})$ amount of space, according to the best known exact algorithms for solving the problem of the global minimum of the function (\ref{E_k}) (known as \textit{the two-way number partitioning problem}, $\mathrm{N}_{\mathrm{PP}}$) \cite{Schroeppel,Ferreira}. At present, it is unknown whether the $\mathrm{N}_{\mathrm{PP}}$ is in the P complexity class\footnote{This might be if P were to be proven to be equal to the NP complexity class; however, it is widely believed that P$\ne$NP \cite{Gasarch}.}, that is, whether an algorithm exists that can solve this problem significantly faster -- in the number of steps upper bounded by a polynomial expression in the size of $N_{\!A}$.\\

\noindent It is clear that in every practical sense a quantity approaching the value of $2^{10^{24}}$ does not differ much from an actual infinity. Hence, even though the quantum states of the Ising model containing a `realistically macroscopic' number $N_{\!A}$ of spins might be effectively calculable, such macroscopic quantum states cannot be calculable \textit{efficiently}, that is, in the amount of time that would \textit{realistically} allow one to find them (and thus to prove explicitly their existence).\\

\section{Conclusion: Non-constructivism and quantum theory}\label{Conclusion}

\noindent The examples shown above demonstrate one thing: By admitting non-constructivist assumptions, quantum theory loses its concrete applicability and thus verifiability of its predictions.\\

\noindent Unlike mathematics that can admit an abstract notion of truth, a physical science must place the emphasis on hands-on provability of existence of any of its theoretical (i.e., mathematical) objects. This means that if two quantum state vectors $\left|\!\left.{\psi_{1}} \!\right.\right\rangle$ and $\left|\!\left.{\psi_{2}} \!\right.\right\rangle$ are asserted to exist, then an explicit example must be given demonstrating the existence of those mathematical objects by outlining an effective procedure (an algorithm) of finding (calculating)  $\left|\!\left.{\psi_{1}} \!\right.\right\rangle$ and $\left|\!\left.{\psi_{2}} \!\right.\right\rangle$. Only in this case a linear combination $\alpha\left|\!\left.{\psi_{1}} \!\right.\right\rangle + \beta\left|\!\left.{\psi_{1}} \!\right.\right\rangle$ (where  $\alpha$ and $\beta$ are complex coefficients satisfying the condition $|\alpha|^2+|\beta|^2=1$) may be constructed and upon that be subject to experimental verifiability.\\

\noindent In contrast, when one assumes a macroscopic quantum superposition and shows that this assumption does not contradict to either basic postulates of the quantum formalism or known experimental data, one still \textit{has not found} the linear combination $\alpha\left|\!\left.{\psi_{1}} \!\right.\right\rangle + \beta\left|\!\left.{\psi_{1}} \!\right.\right\rangle$ and therefore \textit{not proved} its existence.\\

\noindent Furthermore, as it follows from the presented in this paper examples, for at least a few physical systems it is impossible to construct macroscopic quantum state vectors (and hence a superposition of them) for the reason that the existence of such macroscopic quantum states would be based on logical principles allowing one to decide the truth of predicates over \textit{an infinite number of things}.\\

\noindent This permits us to conclude that whereas superpositions of quantum states can be actually found (computed) for microscopic systems, the application of the quantum formalism to any arbitrary macroscopic system cannot be constructively proven.\\


\begin{thebibliography}{16}

\bibitem{Leggett}\label{Leggett}
{A. Leggett}, Testing the limits of quantum mechanics: motivation, state of play, prospects, J. Phys.: Condens. Matter, vol. 14, p. R415–R451, 2002.

\bibitem{Schlosshauer06}\label{Schlosshauer06}
{M. Schlosshauer}, Experimental motivation and empirical consistency in minimal no-collapse quantum mechanics, Ann. Phys., vol. 321, pp. 112-149, 2006.

\bibitem{Gerlich}\label{Gerlich}
{S. Gerlich, S. Eibenberger, M. Tomandl, S. Nimmrichter, K. Hornberger, P. Fagan, J. Tüxen, M. Mayor and M. Arndt}, Quantum interference of large organic molecules, Nature Communications, vol. 2, no. 263, 2011.

\bibitem{Zeilinger}\label{Zeilinger}
{T. Herbst, T. Scheidl, M. Fink, J. Handsteiner, B. Wittmann, R. Ursin and A. Zeilinger}, Teleportation of entanglement over 143 km, February 9, 2015. [Online]. Available: arXiv: arXiv:1403.0009.

\bibitem{Schlosshauer04}\label{Schlosshauer04}
{M. Schlosshauer}, Decoherence, the measurement problem, and interpretations of quantum mechanics, Rev. Mod. Phys., vol. 76, pp. 1267-1305, 2004.

\bibitem{Hajicek}\label{Hajicek}
{ P. Hajicek}, Realism-Completeness-Universality interpretation of quantum mechanics, 18 September 2015. [Online]. Available:  arXiv:1509.05547.

\bibitem{Tammaro}\label{Tammaro}
{E. Tammaro}, Why Current Interpretations of Quantum Mechanics are Deficient, 9 August 2014. [Online]. Available: arXiv:1408.2093.

\bibitem{Bolotin1}\label{Bolotin1}
{A. Bolotin}, Correspondence Principle as Equivalence of Categories, 24 August 2015. [Online]. Available: arXiv:1506.00428.

\bibitem{Landsman}\label{Landsman}
{N. Landsman}, Between classical and quantum, in Handbook of the Philosophy of Science. Vol. 2: Philosophy of Physics, J. Butterfield and J. Earman, Eds., Amsterdam, Elsevier, 2007, pp. 417-554.

\bibitem{Bolotin2}\label{Bolotin2}
{A. Bolotin}, Effectively calculable quantum mechanics, 16 August 2015. [Online]. Available: arXiv:1508.03879.

\bibitem{Altland}\label{Altland}
{A. Altland and B. Simons}, Condensed Matter Field Theory, Cambridge University Press, 2010.

\bibitem{Seroul}\label{Seroul}
{R. Seroul}, Programming for Mathematicians, Berlin: Springer-Verlag, 2000.

\bibitem{Waldschmidt}\label{Waldschmidt}
{M. Waldschmidt}, Diophantine approximation with applications to dynamical systems, in The International Conference on Pure and Applied Mathematics, 2014.

\bibitem{Schroeppel}\label{Schroeppel}
{R. Schroeppel and A. Shamir}, A $T = O(2^{N/2})$, $S = O(2^{N/4})$ algorithm for certain NP-complete problems, SIAM J. Comput., vol. 10, no. 3, 1981.

\bibitem{Ferreira}\label{Ferreira}
{F. Ferreira and J. Fontanari}, Instance Space of the Number Partitioning Problem, Phys. A: Math. Gen., vol. 33, no. 7265, 2000.

\bibitem{Gasarch}\label{Gasarch}
{W. Gasarch}, The P$\stackrel{?}{=}$NP poll, SIGACT News, vol. 33, no. 2, pp. 34-47, 2002.


\end{thebibliography}
\end{document}